\documentclass[11pt,letterpaper]{article}

\usepackage{amsmath}
\usepackage{amsfonts}
\usepackage{amssymb}
\usepackage{graphicx}
\def\be{\begin{equation}}
\def\ee{\end{equation}}

\begin{document}

\begin{flushright} {\footnotesize IC/2006/108}\\ {\footnotesize HUTP-06/A0041}  \end{flushright}
\vspace{5mm}
\vspace{0.5cm}
\begin{center}

\def\thefootnote{\fnsymbol{footnote}}

{\LARGE \bf Limits on $f_{\rm NL}$ parameters from WMAP 3yr data} \\[1cm]
{\Large Paolo Creminelli$^{\rm a}$, Leonardo Senatore$^{\rm b}$, \\
[0.2cm]Matias Zaldarriaga$^{\rm b,c}$ and Max Tegmark$^{\rm d}$}
\\[0.5cm]

{
\textit{$^{\rm a}$ Abdus Salam International Centre for Theoretical
Physics\\ Strada Costiera 11, 34014 Trieste, Italy}} 

\vspace{.2cm}

{ 
\textit{$^{\rm b}$ Jefferson Physical Laboratory, \\
Harvard University, Cambridge, MA 02138, USA}} 

\vspace{.2cm}

{ 
\textit{$^{\rm c}$ Center for Astrophysics, \\
Harvard University, Cambridge, MA 02138, USA
}}

\vspace{.2cm}

{ 
\textit{$^{\rm d}$ Department of Physics, \\
Massachusetts Institute of Technology, Cambridge, MA 02139, USA
}}

\end{center}

\vspace{.8cm}

\hrule \vspace{0.3cm} 
{ \noindent \textbf{Abstract} \\[0.3cm]
\noindent
We analyze the 3-year WMAP data and look for a deviation from
Gaussianity in the form of a 3-point function that has either of the two
theoretically motivated shapes: local and equilateral. There is no
evidence of departure from Gaussianity and the analysis gives the
presently tightest bounds on the parameters $f_{\rm NL}^{\rm local}$ and $f_{\rm NL}^{\rm
  equil.}$, which define the amplitude of respectively the local
and the equilateral non-Gaussianity: $-36 <f_{\rm NL}^{\rm local} <
100$,  $-256 <f_{\rm NL}^{\rm equil.} < 332$ at 95\%
C.L.

\vspace{0.5cm}  \hrule

\def\thefootnote{\arabic{footnote}}
\setcounter{footnote}{0}
\section{Introduction}
During the last few years our understanding of primordial non-Gaussianities of
cosmological perturbations has improved significantly. 

On the theoretical
side it has been firmly established \cite{Maldacena:2002vr,Acquaviva:2002ud} that the simplest models
of inflation sharply predict a level of non-Gaussianity ($\lesssim 10^{-6}$) 
far below the detection threshold of foreseable CMB
experiments. Non-Gaussianity therefore represents a smoking gun for deviations
from this minimal scenario; in fact many alternatives have been studied which give a
much larger non-Gaussianity, an incomplete list including
\cite{Lyth:2002my, Zaldarriaga:2003my,  Boubekeur:2006nj, Creminelli:2003iq,
  Arkani-Hamed:2003uz, Alishahiha:2004eh, Senatore:2004rj,
  Chen:2006nt}. In all these models the deviation from 
pure Gaussian statistics shows up in the 3-point correlator
of density perturbations. This 3-point function contains all the
information about deviation from Gaussianity: no additional
information to constrain these models can be obtained looking for
other signatures of non-Gaussianity, {\em e.g.} Minkowski
functionals or the 4-point function \cite{Creminelli:2006gc}.
Moreover it has been noted that, as every model gives a definite
prediction for the dependence of the 3-point
function on the 3-momenta,  information about the source of 
non-Gaussianity in the early Universe can be recovered by the study of the shape dependence
of the 3-point correlator \cite{Babich:2004gb}. 

From the experimental point of view the progress has been dramatic.
The WMAP experiment has greatly tightened the limits on departures
from Gaussianity. Given the relative simplicity of the physics
describing the CMB, which allows a linearized treatment of
perturbations, and the large data set provided by WMAP, this
experiment alone gives practically all the information we have about
non-Gaussianity nowadays. In \cite{Spergel:2006hy} the recent 3-year
data have been analyzed by the WMAP collaboration; no evidence for
non-Gaussianity has been found and new bounds are obtained. 

The purpose of this paper is to extend this analysis of WMAP 3 year
data, similarly to what was done for the 1 year data
release in \cite{Creminelli:2005hu}. The main difference with respect to the WMAP collaboration
analysis \cite{Spergel:2006hy} is that we look for two different
shapes of the 3-point function. Instead of concentrating only on the
so-called ``local'' shape, we also do the
analysis for the other theoretically motivated shape dependence,
dubbed ``equilateral''. As explained in
\cite{Babich:2004gb,Creminelli:2004yq,Creminelli:2005hu}, it is justified to
concentrate on these two possibilities, both because these are quite
different (so that a single analysis in not good for both) and because
they describe with good approximation all the proposed models producing a high
level of non-Gaussianity. Roughly speaking the local shape is typical 
of multi-field models, while the equilateral one characteristic of single
field models. Given one of the two shapes one puts constraints on the
overall amplitude of the 3-point function, namely on the two parameters
$f_{\rm NL}^{\rm local}$ and $f_{\rm NL}^{\rm equil.}$ \cite{Creminelli:2005hu}. 

Another relevant difference with respect to
\cite{Spergel:2006hy} is the way we cope with the anisotropies of the
noise, which is caused by the fact that some regions of the sky are
observed by the satellite more often than others. We use an improved version of the
estimator as explained in \cite{Creminelli:2005hu}, which allows a 
$\sim 20\%$ tightening of the limits on $f_{\rm NL}^{\rm local}$.

As this paper is an updated version of \cite{Creminelli:2005hu} we
will concentrate (section \ref{sec:diff}) on the new
features of the analysis, referring the reader to the 1st year paper
\cite{Creminelli:2005hu} for all the details. The new limits
on $f_{\rm NL}^{\rm local}$ and $f_{\rm NL}^{\rm equil.}$ are
discussed in section \ref{sec:results} and conclusions are drawn in
section \ref{sec:conclusions}.

\section{\label{sec:diff} Differences with respect to the 1st year analysis}
\subsection{Introduction of the tilt in the analysis.}
As data now favor a deviation from a scale invariant spectrum, the
analysis of the 3-point function has been updated to take into account
the presence of a non-zero tilt. This is completely straightforward in
the case of the local shape; the non-Gaussianity for the Newtonian
potential $\Phi$ is generated by a quadratic term which is local in real space
\be
\Phi({\bf x}) = \Phi_g({\bf x}) + f_{\rm NL}^{\rm local} (\Phi_g^2({\bf
x})- \langle\Phi_g^2\rangle) \;,
\ee
where $\Phi_g$ is a Gaussian variable.   
In the presence of a non-zero tilt this gives in Fourier space
\be
 \langle \Phi({\bf k}_1) \Phi({\bf k}_2) \Phi({\bf k}_3) \rangle = (2 \pi)^3 \delta^3
\big({\bf k}_1 + {\bf k}_2 + {\bf k}_3 \big) F( k_1,  k_2 ,  k_3) 
\ee
with
\begin{eqnarray}
\label{eq:local}
F( k_1,  k_2 ,  k_3) & = & 2 f_{\rm
  NL}^{\rm local} \cdot \left[P(k_1) P(k_2) +P(k_1) P(k_3) + P(k_2) P(k_3)
\right] \\ & = & f_{\rm NL}^{\rm local} \cdot 2 \Delta_\Phi^2
\cdot \left(\frac1{k_1^{3-(n_s-1)} k_2^{3-(n_s-1)}} + \frac1{k_1^{3-(n_s-1)} k_3^{3-(n_s-1)}} + 
\frac1{k_2^{3-(n_s-1)} k_3^{3-(n_s-1)}}\right)\nonumber
\end{eqnarray}
where $P(k)$ is the power spectrum $\langle \Phi({\bf k}_1) \Phi({\bf k}_2)
\rangle \equiv (2\pi)^3 \delta^3\big({\bf k}_1 + {\bf k}_2\big) P(k_1)$, with normalization $\Delta_\Phi$  and
tilt $n_s$: $P(k) \equiv \Delta_\Phi \cdot k^{-3+(n_s-1)}$.
We remind the reader that the function $F$ enters in the estimator as:
\begin{eqnarray}
\label{eq:explestim}
{\cal{E}} & = & \frac1N \cdot \sum_{l_i m_i} \int 
d^2\hat{n} \; Y_{l_1 m_1}(\hat{n}) Y_{l_2 m_2}(\hat{n})Y_{l_3 m_3}(\hat{n}) 
\int \limits^{\infty}_0 r^2 dr \; j_{l_1}(k_1r) j_{l_2}(k_2r) j_{l_3}(k_3r) \; C_{l_1}^{-1} C_{l_2}^{-1} C_{l_3}^{-1} 
\nonumber \\ & & \int \frac{2 k^2_1 dk_1}{\pi} \frac{2 k^2_2 dk_2}{\pi} \frac{2 k^2_3 dk_3}{\pi} 
F(k_1,k_2,k_3) \Delta^T_{l_1}(k_1)
\Delta^T_{l_2}(k_2) \Delta^T_{l_3}(k_3) \; a_{l_1 m_1}a_{l_2 m_2}a_{l_3 m_3} \;,
\end{eqnarray}
where $\Delta^T_l(k)$ is the CMB transfer function, and where for simplicity we have neglected the term 
linear in the data discussed in \cite{Creminelli:2005hu}.
It is straightforward to check that the only modification introduced
in the analysis is that the function $\beta_l(r)$ defined
in \cite{Komatsu:2003iq} now depends on the tilt, while the function $\alpha_l(r)$
remains unchanged:
\begin{eqnarray}
\alpha_l(r) & \equiv & \frac2\pi \int_0^{+ \infty} \!\!\! dk \; k^2 \, \Delta^T_l(k) j_l(k r) \\
\beta_l(r) & \equiv & \frac2\pi \int_0^{+ \infty} \!\!\! dk \;
k^{-1+(n_s-1)} \, \Delta^T_l(k) j_l(k r)\Delta_\Phi\;.
\end{eqnarray}
As discussed in \cite{Creminelli:2005hu} the analysis of the
equilateral shape is done using a template function $F(k_1,k_2,k_3)$, which is very
similar (with few percent corrections) to the different shapes
predicted by equilateral models, and at the same time sufficiently
simple to make the analysis feasible. For the equilateral case the way to take into account a non-zero
tilt is not unique. In models which predict this shape of
non-Gaussianity the evolution with scale of the 3-point function
is not fixed by the tilt of the spectrum\footnote{What {\em is} fixed
by the spectrum is the squeezed limit of any single field model \cite{Maldacena:2002vr,Creminelli:2004yq}.
Without any slow-roll approximation
 \be
 \label{eq:maldacenalimit}
\lim_{k_1 \to 0} \langle\zeta_{\vec k_1} \zeta_{\vec k_2} \zeta_{\vec
   k_3}\rangle = - (2 \pi)^3 \delta^3(\sum_i \vec k_i) P_{k_1} P_{k_3}
 \frac{d \log k_3^3 P_{k_3}}{d \log k_3} \;. \ee
This just tells us that the signal of non-Gaussianity is very small in the
squeezed limit, but it does not help in fixing the scale dependence of
the 3-point function for configurations close to equilateral, where the signal is
concentrated.
}. 

For consistency with the
local shape we can define new $\gamma_l(r)$ and $\delta_l(r)$
functions as
\begin{eqnarray}
\gamma_l(r) & \equiv & \frac2\pi \int_0^{+ \infty} \!\!\! dk \;
k^{1+\frac13 (n_s-1)} \, \Delta^T_l(k) j_l(k r) \Delta_\Phi^{1/3}\\
\delta_l(r) & \equiv & \frac2\pi \int_0^{+ \infty} \!\!\! dk \;
k^{\frac23 (n_s-1)} \, \Delta^T_l(k) j_l(k r)\Delta_\Phi^{2/3} \;.
\end{eqnarray}

The new template shape will be of the form\footnote{Notice that the
  relationship between the new and the old, scale invariant, template
  function used in \cite{Creminelli:2005hu} is simply $F_{\rm new}(k_1,k_2,k_3) = F_{\rm old}
  (k_1^{1-\frac13 (n_s-1)},k_2^{1-\frac13 (n_s-1)},k_3^{1-\frac13
    (n_s-1)})$. From this we see that the function goes as
  $k_1^{-1+\frac13(n_s-1)}$ for $k_1 \to 0$, while in the local case the
signal is much more enhanced going as $k_1^{-3+(n_s-1)}$.}
\begin{eqnarray}
\label{eq:ours}
\nonumber F & = &  f_{\rm NL}^{\rm equil.} \cdot 6  
 \Delta_\Phi^2 \cdot \left(-\frac1{k_1^{3-(n_s-1)} k_2^{3-(n_s-1)}} +
   (2 \; perm.) - \frac2{(k_1 k_2 k_3)^{2-\frac23(n_s-1)} } \right. \\ & & \left. +
   \frac1{k_1^{1-\frac13(n_s-1)} k_2^{2-\frac23(n_s-1)}  k_3^{3-(n_s-1)}}
+ (5 \; perm.) \right) \;.
\end{eqnarray}
As discussed above we expect that the difference between a given model
and this template shape, taking also into account differences in the evolution with
scale, to be small. Until a clear detection of non-Gaussianity is
found, the use of a single template shape for the whole class of
``equilateral  models'' is justified.

\subsection{Improved combination of the maps.}
In the non-Gaussianity analysis \cite{Komatsu:2003fd,Spergel:2006hy} and
\cite{Creminelli:2005hu} the 8 maps at different frequency Q1, Q2, V1,
V2, W1, W2, W3 and W4 were combined making a pixel by pixel average
weighted by the noise $\sigma_0^2/N_{\rm obs}$, where $N_{\rm obs}$ is
the number of observations of the pixel and $\sigma_0$ is a band
dependent constant. As the $N_{\rm obs}$ maps are very similar to each
other, the procedure amounts to take a pixel-independent combination 
of the maps, weighted by the average noise. This procedure however is
not really optimal because it neglects the effect of the beams: at
high $l$ one should give more weight to the $W$ bands as they have the
narrowest beam. In other words the optimal procedure is an
$l$-dependent combination with signal-to-noise weight \cite{Tegmark:2003ve}. The difference
with the naive combination becomes more and more relevant going to
higher multipoles, when the effect of the beams is relevant. Thus the
improvement becomes more important as time passes and noise
reduces, allowing to explore regions of higher $l$. 
Making a combination in Fourier space has however some disadvantage: given the
non-locality of the combination in real space it is not clear how to
proceed in masking the contaminated regions of the sky.  

We choose to use an intermediate procedure. We combine the
maps with a constant coefficient as in the previous analyses, but
instead of using the noise as weight, we use a weight based on the  
signal-to-noise ratio at a given fixed multipole $l_{\rm comb}$. If we
choose a very small $l_{\rm comb}$, the signal is the same in all the maps and we
are back to a noise weight, which is optimal for the lowest
multipoles. 
On the other hand if $l_{\rm comb}$ is
large we are making an optimal combination at high $l$, but not so
good at low $l$. We tried many values of $l_{\rm comb}$ to get a
result which is as close as possible to the optimal combination. The
choice $l_{\rm comb}=235$ turns out to be the best one. We estimated
that it improves the limits on the $f_{\rm NL}$ parameters by $\sim
1.4 \%$ with respect to the original noise weighting. A very small
further improvement, of order $0.4 \%$, could be theoretically achieved with
the optimal signal-to-noise weighting. Given the problems in dealing
with the masking of foregrounds, it is clearly not worthwhile using an
$l$-dependent combination. In figure (\ref{fig:comb}) we
compare the different ways of combining the maps, showing the
effective noise as a function of $l$, compared with the signal. We see
that $l_{\rm comb}=235$ gives something very close to optimal.

\begin{figure}[ht!]             
\begin{center}
\includegraphics[width=14.0cm]{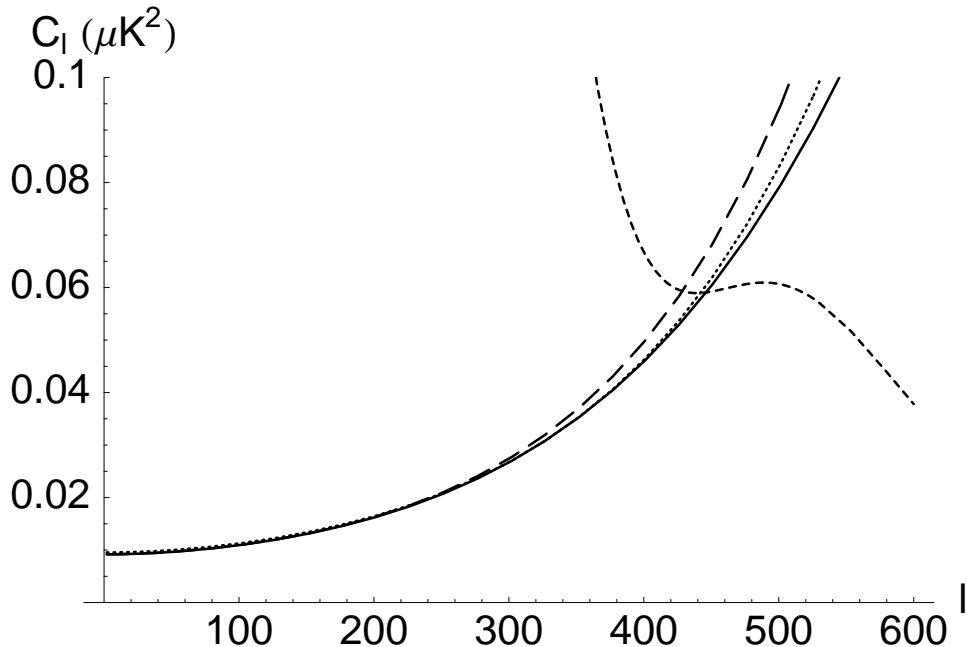}\vspace{0.8cm}
\caption{\label{fig:comb} \normalsize Effective noise for different
  combinations of the maps compared with the signal
  (short-dashed). Optimal signal-to-noise combination (solid); the
  combination used in the analysis, with $l_{\rm comb} =235$ (dotted); and
  the noise weighted combination, equivalent to $l_{\rm comb} = 0$ (long-dashed)
  used in the 1st year analysis \cite{Creminelli:2005hu} and in the
  3yr analysis by the WMAP collaboration \cite{Spergel:2006hy}.}
\end{center}
\end{figure}

\subsection{Variation of the cosmological parameters.} We use for our analysis
the Power Law $\Lambda$CDM Model which gives the best fit to WMAP 3 year
data (see Table 2 of \cite{Spergel:2006hy}). The cosmological
parameters are given by: $\Omega_b h^2 = 0.0223$, $\Omega_m h^2 = 0.128$, $h =0.73$, $\tau = 0.092$, 
$n_s =0.958$. The most notable differences with respect to the first year parameters
are the drop in the optical depth and the presence of a non-zero tilt
of the spectrum. 

Let us try to estimate the effect of these changes on the variance of
the estimators for the $f_{\rm NL}$ parameters. 
Roughly speaking the reionization optical depth $\tau$ enters as a multiplicative factor
$e^{-\tau}$ in front of the transfer function $\Delta_l^T(k)$ for $l$
corresponding to scales shorter than the horizon at reionization. In first approximation one
can assume that the points on the first peak remain unchanged: as their error is very small the normalization will
change to keep the power unchanged there. 
This means that the reduction of the best fit value for $\tau$ from $\tau=0.17$
to $\tau=0.092$ will be compensated by a decrease in the amplitude of
perturbations: $\Delta_\Phi$ will decrease by approximately
$16\%$. The level of non-Gaussianity of the fluctuations is given by $f_{\rm NL} \cdot
\Delta_\Phi^{1/2}$, so that a decrease in the amplitude of
perturbation will relax the constraints on the $f_{\rm NL}$
parameters. The decrease in the optical depth enlarges the error on
the $f_{\rm NL}$ parameters by $\sim 8 \%$. 

Let us now discuss the effect of the tilt. We can schematize the
addition of a red tilt as a reduction of the power $\Delta_{\rm
  short}$ at scales shorter than the first peak and an enhancement for
larger scales ($\Delta_{\rm long}$). What is the effect of this on the
limits we can put on the $f_{\rm NL}$ parameters? For the local shape
the signal is mostly coming from squeezed triangles in Fourier space,
with one side much shorter than the others \cite{Babich:2004gb}. The non-Gaussian signal
for these triangles goes as $\Delta_{\rm long} \cdot \Delta_{\rm
  short}$ (see eq.~(\ref{eq:local})), while the error, {\em i.e.}~the
typical value in a realization with pure Gaussian statistics, will go as $\Delta_{\rm
  long}^{1/2} \cdot \Delta_{\rm short}^{1/2} \cdot  \Delta_{\rm
  short}^{1/2}$. Thus the limits on $f_{\rm NL}^{\rm local}$ will
become tighter proportionally to $\Delta_{\rm long}^{1/2}$. For the
equilateral shape the signal is coming from equilateral
configurations. The contribution from triangles with $l$ smaller than
the first peak will have more signal while triangles on
shorter scale will have less. There is a mild
cancellation between these effects giving a rather small effect in the
equilateral case. One can check this intuition with a numerical
analysis of the variance of the estimator varying the tilt. Going from
a flat spectrum to the value of $n_s$ favored by WMAP 3yr data, the 
constraint on $f_{\rm NL}^{\rm local}$ becomes tighter by 9\%, while
the one on $f_{\rm NL}^{\rm equil.}$ becomes looser by 5\%.

Obviously this way of taking into account the variation in the
knowledge of the cosmological parameters is quite naive. One should
properly marginalize over all parameters when quoting the final limits
on the non-Gaussianity. This would slightly enlarge the allowed range,
roughly by an amount comparable to the variation induced by the
change of the cosmological parameters discussed above. This approach
is numerically very demanding but it
will be mandatory if a significant detection of non-Gaussianity will be achieved. 

One can put together the variations induced by the change in the
cosmological parameters with the reduction of the noise given the
additional amount of data, to estimate
the final improvement on the allowed range for $f_{\rm NL}$. With a
scale invariant signal the noise
reduction would give a $\sqrt{3}$ improvement. But this is not a good
approximation, both because the transfer function imprints features on
the scale invariant primordial spectrum and because the beams
cut off the signal at high $l$. A good estimate of the improvement is to look at the
multipole where signal and noise intersect, with 1st and 3yr data. The
constraints on the non-Gaussianity parameters will scale as
$N_{\rm pixels}^{-1/2} \propto l_{\rm max}^{-1}$. The multipole of
intersection increases by $\sim 20$\%, so that we naively expect a $\sim 20$\% reduction of the limits.
If one puts this together with the discussion above a $\sim 20$\%
improvement for $f_{\rm NL}^{\rm local}$ is expected, while the improvement on
$f_{\rm NL}^{\rm equil.}$ will be really marginal.

\section{\label{sec:results} Results of the analysis}
Aside from the few differences discussed above, the analysis of the 3-year
data strictly follows what was done for the 1st year release in
\cite{Creminelli:2005hu}. 

For the local shape analysis, the inhomogeneity of the noise, which reflects
the fact that different regions of the sky are observed a different
number of times, causes some trouble when one extends the analysis to
high $l$ where the noise is relevant. The variance of the naive
trilinear estimator, which would be optimal in the presence of
rotational invariance, starts increasing at a certain point while
including more and more data at short scale
(see figure \ref{fig:local}). 
This was already noted by the WMAP collaboration in
\cite{Komatsu:2003fd}. A partial solution of the problem was given in
\cite{Creminelli:2005hu}, with the introduction of an improved
estimator. The improvement simply consists in an additional term
which is linear in the multipoles $a_{lm}$. We stress that this
additional term by construction vanishes on average, independently of
the value of $f_{\rm NL}^{\rm local}$. In other words it does not bias
the estimator, just reduces its variance. 
In figure \ref{fig:local} we see that the behavior at high $l$ of the
new estimator is greatly improved, with a resulting $\sim 20$\% improvement on the limits on
$f_{\rm NL}^{\rm local}$. The estimator has minimum variance for
$l_{\rm max}=370$, with a standard deviation of $34$. Unfortunately
this is not a big improvement with respect to the 1st year analysis which
had a standard deviation of $37$. From figure \ref{fig:local} we see that an optimal analysis, which would require the full inversion of the covariance matrix, should be able to further reduce the standard deviation to $25$, a $\sim 25$\% improvement with respect to our result.

\begin{figure}[ht!]             
\begin{center}
\includegraphics[width=13.0cm]{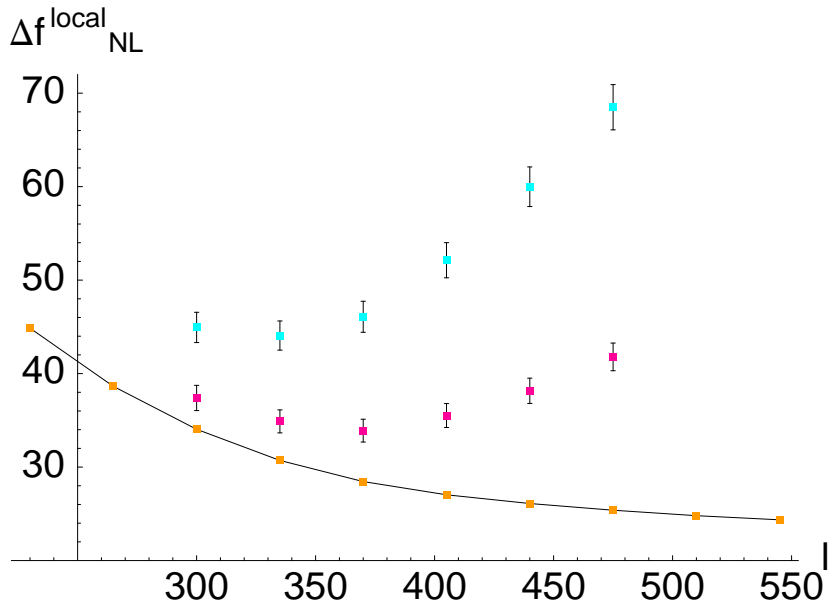}\vspace{0.8cm}
\caption{\label{fig:local} Standard deviation for estimators of $f_{\rm NL}^{\rm local}$ as a 
function of the maximum $l$ used in the analysis.  The combination of
the maps is done using $l_{\rm comb} = 235$.
Lower curve: lower bound deduced from the full sky variance. Lower data points: standard deviation for the
trilinear + linear estimator (see details in \cite{Creminelli:2005hu}). Upper data points:
the same for the estimator without linear term used in
\cite{Spergel:2006hy}.  The error bars are
not independent as the results at different $l$ are all based on the
same set of MonteCarlo maps.}
\end{center}
\end{figure}

We applied the estimator to the foreground reduced WMAP sky maps data (see \cite{Hinshaw:2006ia} for a discussion about foreground removal)  and we obtain $f_{\rm NL}^{\rm
  local}= 32$. There is therefore no evidence of deviation from a
Gaussian statistics \footnote{We notice that if we apply our estimator to the WMAP sky 
maps data without foreground subtraction, we still see no evidence of deviation from 
Gaussianity both in the case of $f_{\rm NL}^{\rm
  local}$ and of $f_{\rm NL}^{\rm
  equil.}$.}. The new limits on  $f_{\rm NL}^{\rm local}$ are:
\be
\label{eq:locfinal}
-36 <f_{\rm NL}^{\rm local} < 100 \quad {\rm at} \;95\% \;{\rm C.L.}
\ee

As explained in \cite{Creminelli:2005hu} the effect of the noise
inhomogeneities is small for the equilateral shape, so that any
improvement of the estimator is useless (see figure
\ref{fig:equil}). Following the discussion in the section above, it is
easy to understand that the red tilt makes the curve in figure
\ref{fig:equil} flatten out faster than in the local case (figure
\ref{fig:local}): the contribution at high $l$ is suppressed. The estimator does not  get worse going to higher
$l$ and we can do the analysis at $l_{\rm max} =475$. The standard deviation of
the estimator is $147$. As expected the improvement with respect to
the first year analysis is marginal, only $\sim 3\%$. The value obtained on the real maps is $f_{\rm NL}^{\rm equil.} =38$. Again there is no evidence of non-Gaussianity of this shape and the new allowed range is given by
\be
\label{eq:eqfinal}
-256 <f_{\rm NL}^{\rm equil.} < 332 \quad {\rm at} \;95\% \;{\rm C.L.}
\ee    
An optimal analysis should be able to reduce the range by $\sim 15$\% (see figure \ref{fig:equil}).

\begin{figure}[ht!]             
\begin{center}
\includegraphics[width=12.5cm]{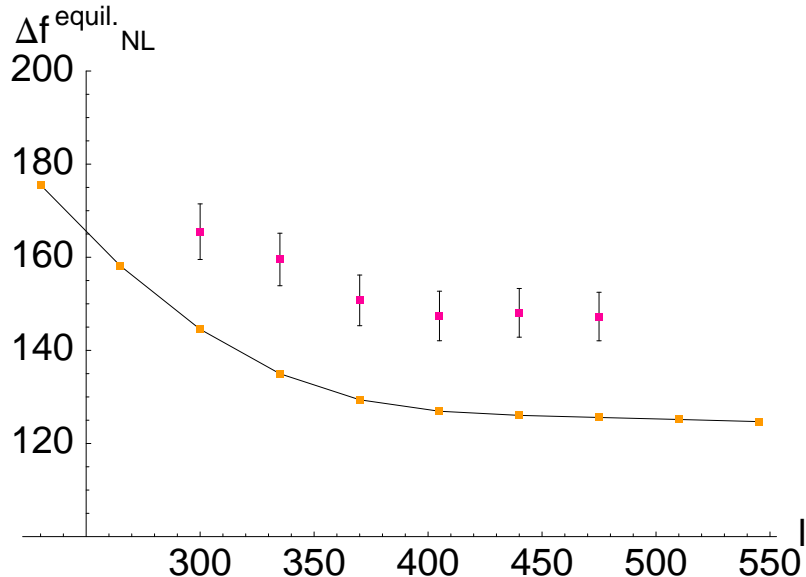}\vspace{0.8cm}
\caption{\label{fig:equil} Standard deviation of the estimator of $f_{\rm NL}^{\rm equil.}$ as a 
function of the maximum $l$ used in the analysis. The combination of
the maps is done using $l_{\rm comb} = 235$.
Lower curve: lower bound deduced from the full sky variance. Data points: standard deviation for the
estimator used in the analysis.  The error bars are
not independent as the results at different $l$ are all based on the
same set of MonteCarlo maps.}
\end{center}
\end{figure}

\section{\label{sec:conclusions}Conclusions}
We have analyzed the 3-year WMAP data and looked for theoretically motivated deviations from
Gaussianity. Inflationary models which give an observable amount of
non-Gaussianity predict a 3-point function of either local or
equilateral form. In the data there is no evidence of deviation from
Gaussianity, so that we derive bounds on the parameters $f_{\rm
  NL}^{\rm local}$ and $f_{\rm NL}^{\rm equil.}$, the amplitude of the local and
equilateral shape, respectively. The results are 
\begin{eqnarray}
-36 < & f_{\rm NL}^{\rm local} & < 100 \quad {\rm at} \;95\% \;{\rm C.L.} \\
-256 <& f_{\rm NL}^{\rm equil.} & < 332 \quad {\rm at} \;95\% \;{\rm C.L.}
\end{eqnarray}
The improvement with respect to the 1st year analysis is marginal
($\sim 10\%$ in the local case and $\sim 3\%$ for the equilateral
case): the expected $\sim 20\%$ improvement from noise reduction is
partially compensated by a change in the best fit cosmological
parameters. A further $\sim 20\%$ improvement is expected with 8-year
statistics. In the future, this kind of analysis will allow to extract in a numerically
feasible way almost all the information about $f_{\rm
  NL}^{\rm local}$ and $f_{\rm NL}^{\rm equil.}$ from Planck data.
The signal should be dominant until $l \sim 1500$, so
that one expects a factor of 4 improvement. In addition, polarization
measurements can further enhance the sensitivity by a factor of 1.6 \cite{Babich:2004yc}.

We stress that our approach is to look only for forms of non-Gaussianity which are theoretically
motivated within the inflationary paradigm. Our results are not in contradiction
with some claims in the literature of detection of a non-Gaussian statistics different from ours 
(see for example \cite{McEwen:2006yc,Land:2006bn} and references therein). Given the fact there are infinite ways
of deviating from Gaussiannity, it remains difficult to assess the statistical significance of those
results.

\section*{Acknowledgments}
We would like to thank Eiichiro Komatsu and Alberto Nicolis for their help during the project, 
Daniel Babich and Suvendra Nath Dutta for help with analysis software, and Liam McAllister for 
pointing out an error in the first version of this paper.
The numerical analysis necessary for the completion of this paper was performed on the Sauron cluster, 
at the Center for Astrophysics, Harvard University, making extensive use of the HEALPix package \cite{Gorski:2004by}. 
MZ was supported by the Packard and Sloan foundations, NSF AST-0506556 
and NASA NNG05GG84G. MT was supported by NASA grant NNG06GC55G, NSF
grants AST-0134999 and 0607597, the Kavli 
Foundation and the Packard Foundation.

\end{document}